\documentclass[9pt,twocolumn,twoside]{opticajnl}
\journal{opticajournal} % use for journal or Optica Open submissions

\setboolean{shortarticle}{true}
\usepackage{lineno}
\title{Robust multimode interference and conversion in topological unidirectional surface magnetoplasmons}

\author[1,2]{Chao Liu}
\author[2]{Ziyang Zhao}
\author[2]{Tianjing Guo}
\author[3]{Jie Xu}
\author[2]{Xiaohua Deng}
\author[1]{Kai Yuan}
\author[2]{Rongxin Tang}
\author[4,5]{Kosmas L. Tsakmakidis}
\author[1,2,*]{Lujun Hong}

\affil[1]{School of Information Engineering, Nanchang University, Nanchang 330031, China}
\affil[2]{Institute of Space Science and Technology, Nanchang University, Nanchang 330031, China}
\affil[3]{School of Medical Information and Engineering, Southwest Medical University, Luzhou 646000, China}
\affil[4]{Section of Condensed Matter, Physics Department of Physics, National and Kapodistrian University of Athens Panepistimioupolis, Athens GR-157 84, Greece}
\affil[5]{ktsakmakidis@phys.uoa.gr}
\affil[*]{ljhong@ncu.edu.cn}

\begin{abstract}
	We have theoretically investigated surface magnetoplasmons (SMPs) in a yttrium-iron-garnet (YIG) sandwiched waveguide. The dispersion demonstated that this waveguide can support topological unidirectional SMPs. Based on unidirectional SMPs, magnetically controllable multimode interference (MMI) is verified in both symmetric and asymmetric waveguides. Due to the coupling between the modes along two YIG-air interfaces, the asymmetric waveguide supports a unidirectional even mode within a single-mode frequency range. Moreover, these modes are topological protected when disorder is introduced. Utilizing robust unidirectional SMPs MMI (USMMI), tunable splitters have been achieved. It has been demonstrated that mode conversion between different modes can be realized. These results provide many degrees of freedom to manipulate topological waves.
	
\end{abstract}

\setboolean{displaycopyright}{false} % Do not include copyright or licensing information in submission.

\begin{document}
	
	\maketitle
	
	\section{Introduction}
	Topological unidirectional waves have attracted much attention due to their unique optical properties of wave propagation protected from backscattering \cite{lu2014NP,cheng2016NM,ozawa2019RMP,chen2022PRL}. As analogs of quantum Hall edge states in photonic crystals (PhCs) \cite{haldane2008PRL}, unidirectional edge modes were proven in YIG PhCs \cite{wang2008PRL}, and they were first experimentally observed at microwave frequencies \cite{wang2009NATURE}. Due to the time-reversal symmetry broken by external magnetic field (EMF), such modes can travel in only one direction and are robust against backscattering from disorder \cite{wang2021PRL,poo2011PRL}. As another type of unidirectional mode, surface magnetoplasmons (SMPs) were also proposed \cite{yu2008PRL,gao2016NC}, attracting great interest due to the rich physics of nonreciprocal and topological materials \cite{jin2017PRL,tsakmakidis2017SCIENCE,hong2021OME,liang2021Optica}. Recently, topologically unidirectional SMP  propagation was experimentally verified in a YIG-based SMP waveguide \cite{li2024NC}.
	
	Owing to their nontrivial topologically protected properties \cite{Gao2023NC,jinT2016NC}, unidirectional modes based on PhCs or SMPs are suitable for realizing topologically optical devices, such as logic gates \cite{xu2023AOM}, lasers \cite{bahari2017science}, slow light \cite{mann2021PRL}, and splitters \cite{tee2014OE,chen2021PRL}. Recently, multimode interference (MMI) was achieved using topological PhCs, demonstrating robustness against disorder \cite{liu2022OL, liu2023OLT}. Mode conversion has also been realized in a YIG-based PhCs waveguide \cite{huang2023OL}. More recently, magnetically controllable MMI based on topological YIG-PhCs was demonstrated \cite{tang2024LSA}. It is a natural desire to investigate whether an SMP waveguide can achieve MMI and mode conversion. In this Letter, we will show that magnetically controllable unidirectional SMPs MMI (USMMI) can be achieved. Based on such USMMI, tunable splitters are designed in symmetric and asymmetric structures, demonstrating robustness against disorder. Notably, a unidirectional even mode occurs within a single-mode frequency range in the asymmetric waveguide, unlike in conventional SMP waveguides. This finding enables us to achieve efficient mode conversion through the coupling of different waveguides. %Moreover, mode conversion can be achieved with the insertion of metal.

\begin{figure}[t]
		\centering\includegraphics[width=3 in]{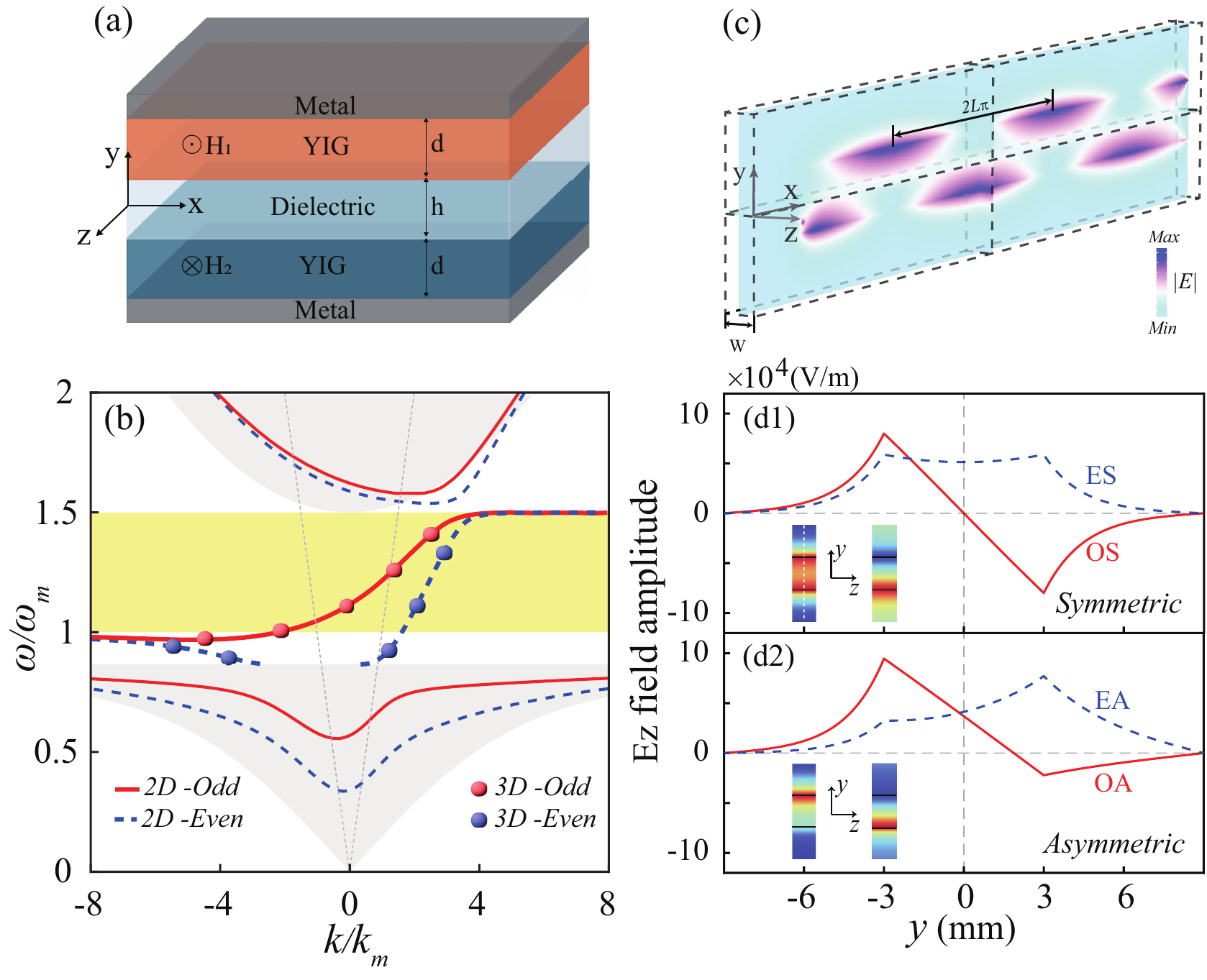}
		\caption{(a)  Schematic of the proposed topological waveguide with opposing EMFs in two YIG slabs. (b) Dispersion relation of odd mode (solid lines) and even mode (dashed lines) in the 2D symmetric structure. Circles indicate results for the 3D realistic system. The unidirectional propagation occurs in $[\omega_{m},1.5\omega_m]$, marked by the yellow. The gray shaded area represents the YIG bulk modes. (c) Simulated $E$-field amplitude at $\omega=1.1\omega_m$ in the  3D waveguide. Distribution of $E_z$ field along the y-axis in symmetric (d1) and asymmetric (d2) structures. Insets show the mode profiles. The parameters are: $d=0.1\lambda_m$, $h=0.1\lambda_m$, $W=0.05\lambda_m$, and $H_1=H_2=893$ G.}
		\label{fig1}
	\end{figure}

	We consider a waveguide composed of two YIG slabs sandwiched between metal and dielectric, as shown in Fig. \ref{fig1}(a). The dielectric layer with thickness $h$ has a permittivity of ${\varepsilon _r}$. The two YIG slabs with thickness $d$ are magnetized by two opposing EMFs ($H_1$ and $H_2$), along the $\pm z$ direction. Owing to the EMFs, the YIG slabs are gyromagnetically anisotropic with relative permittivity of ${\varepsilon _m=15}$ and permeability tensor ${\mu _m}$ \cite{fang2024OLT,chen2024PRL}.
	\begin{eqnarray}
		{\mu}^{+}_m = \left[
		{\begin{array}{*{20}{c}}
				{{\mu _{1}}}&{{-i\mu_2}}&0\\
				{{  i{\mu_2}}}&{{\mu_1}}&0\\
				0&0&1
		\end{array}} \right],\quad
		{\mu}^{-}_m = \left[
		{\begin{array}{*{20}{c}}
				{{\mu' _{1}}}&{{ i\mu'_2}}&0\\
				{{-  i{\mu'_2}}}&{{\mu'_1}}&0\\
				0&0&1
		\end{array}} \right]
	\end{eqnarray}
	with $\mu_1$= $1+\dfrac{\omega_m ( \omega_0 + i\alpha \omega)}{(\omega_0+i\alpha\omega)^2 - \omega^2}$, $\mu_2$= $\dfrac{\omega_m \omega}{(\omega_0+i\alpha\omega)^2 - \omega^2}$, $\mu_1'$= $1+\dfrac{\omega_m ( \omega_0' + i\alpha \omega)}{(\omega_0'+i\alpha\omega)^2 - \omega^2}$, and $\mu_2'$ = $\dfrac{\omega_m \omega}{(\omega_0'+i\alpha\omega)^2 - \omega^2}$, where $\omega_0$= $2\pi \gamma H_1$, $\omega_0'$= $2\pi \gamma H_2$ (${\gamma}=2.8$~MHz/G is the gyromagnetic ratio) is resonance frequency, ${\omega}$ is the angular frequency, $\alpha$ is damping coefficient, and ${\omega _m}$ is the characteristic circular frequency. This waveguide can support the transverse electric (TE) mode ($H_x, H_y, E_z$). By solving Maxwell's equations with the continuous boundary conditions, the dispersion relation of SMPs can be derived analytically as (see the details in Supplement 1)
	\begin{equation}
		e^{2a_rh}=\dfrac{\left(1-\dfrac{M}{\alpha_r\mu_v}\right)
			\left(1-\dfrac{N}{\alpha_r\mu_v'}\right)}
		{\left(1+\dfrac{M}{\alpha_r\mu_v}\right)
			\left(1+\dfrac{N}{\alpha_r\mu_v'}\right)}
		\label{eq:eq2}
	\end{equation}
with $M=k\dfrac{\mu_2}{\mu_1} + \dfrac{\alpha_1}{\tanh{\alpha_1d}}$ and $N=k\dfrac{\mu'_2}{\mu'_1} + \dfrac{\alpha_2}{\tanh{\alpha_2d}}$, where $k$ is the propagation constant, $\alpha_{r}=\sqrt{k^{2}-\epsilon_{r}k_{0}^{2}}$ ($k_0=\omega/c$ is the vacuum wavenumber), $\alpha_{1}=\sqrt{k^{2} -\mu_{\rm{v}}\epsilon_{m}k_{0}^{2}}$, $\alpha_{2}=\sqrt{k^{2} -\mu'_{\rm{v}}\epsilon_{m}k_{0}^{2}}$ ($\mu_{\rm{v}}=\mu_1-\mu_{2}^2/\mu_1$ and ${\mu'_{\rm{v}}}={\mu'_1}-{\mu'_{2}}^2/{\mu'_1}$ are the Voigt permeabilities) are the attenuation coefficients in the dielectric, upper and lower YIG slabs, respectively. It is found from \eqref{eq:eq2} that SMPs have four asymptotic frequencies when $k\rightarrow \pm\infty $: $\omega_{\rm{sp1}}=\omega_{0}+0.5\omega_m$, $\omega_{\rm{sp2}}=\omega_{0}+\omega_m$, $\omega_{\rm{sp3}}=\omega'_{0}+0.5\omega_m$, and $\omega_{\rm{sp4}}=\omega'_{0}+\omega_m$. In the special case of ${H_1} = {H_2}$, the dispersion relation of SMPs in \eqref{eq:eq2} can be simplified to
	\begin{subequations}
		\begin{eqnarray}
			k\dfrac{\mu_2}{\mu_1} + \dfrac{\alpha_1}{\tanh{\alpha_1d}} +\alpha _{r}\mu_{\rm{v}}{\rm{tanh}}\left (\dfrac{\alpha _{r}h}{2} \right )=0 \quad(ES) \label{eq:eq3a}\\
			k\dfrac{\mu_2}{\mu_1} + \dfrac{\alpha_1}{\tanh{\alpha_1d}}+
			\alpha _{r}\mu_{\rm{v}}{\rm{coth}}\left (\dfrac{\alpha _{r}h}{2} \right )=0 \quad(OS)
			\label{eq:eq3b}
		\end{eqnarray}
		\label{eq:eq3}
	\end{subequations}
	for the even-symmetric (ES) and odd-symmetric (OS) modes, respectively. The presence of the linear term \( k \) in \eqref{eq:eq3} leads to different dispersion for forward and backward propagation, resulting in non-reciprocity.

	%\subsection{symmetric structure}
	\begin{figure}[t!]
		\centering\includegraphics[width=2.8 in]{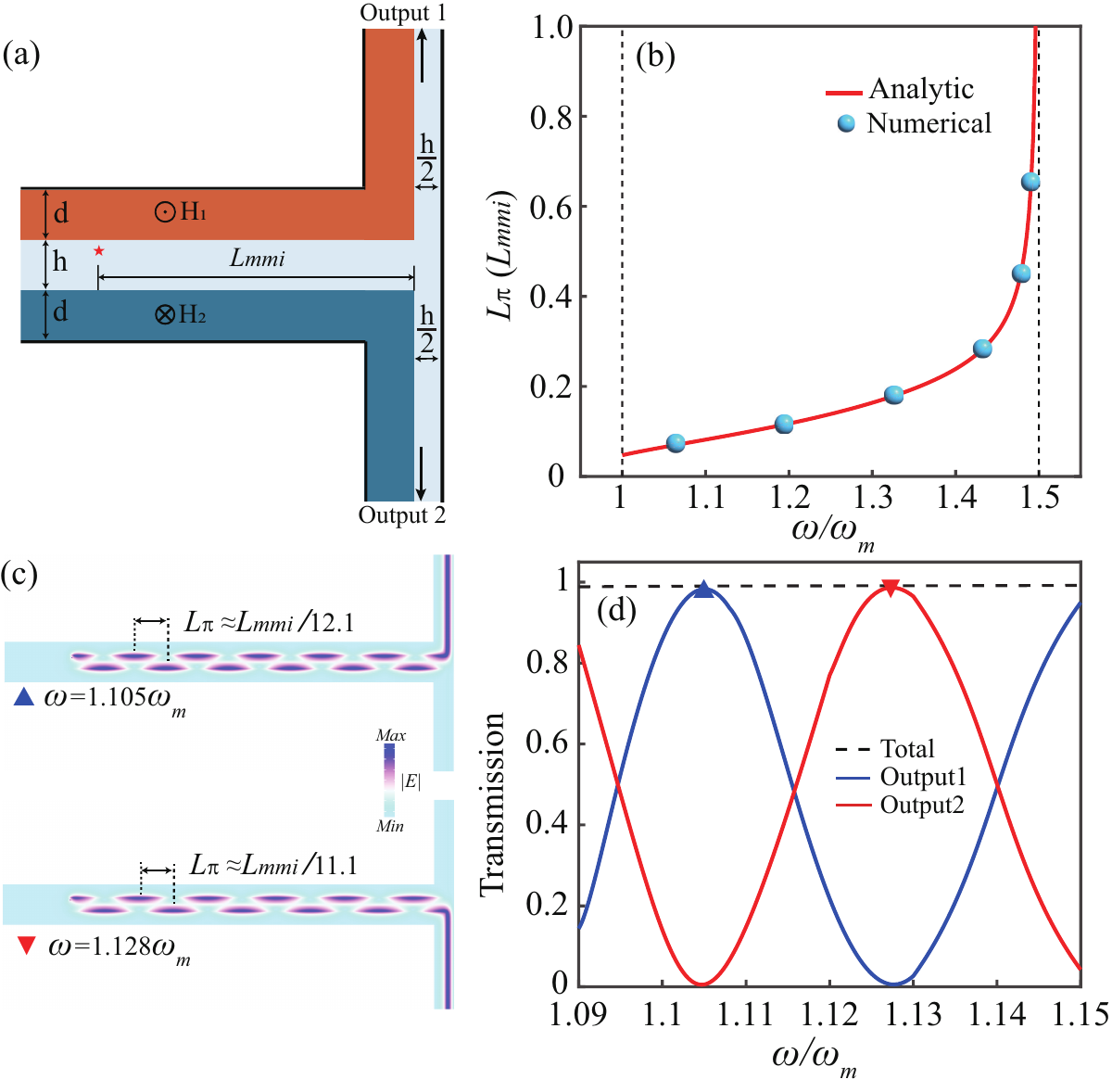}
		\caption{ (a) Schematic of the splitter based on USMMI. (b) Analytical (solid line) and numerical (circles) results of beat length $L_\pi$ as a function of $\omega$. (c) Simulated $E$-field amplitues in the symmetric splitter at $\omega =1.105 \omega_{m}$ and  $1.128\omega_{m}$. (d) Transmission coefficients of the symmetric splitter ($H_1=H_2=893$ G) as a function of $\omega$.}\label{fig2}
	\end{figure}
	First, we consider a symmetric structure ($H_1 = H_2$). The dispersion of SMPs in this waveguide can be numerically calculated using \eqref{eq:eq3}. Here, we take $d = 0.1 \lambda_\mathrm{m}$ ($\lambda_\mathrm{m}$=$2\pi c/\omega_{m}$), and ${\omega_m = 10\pi \times 10^{9}}$~rad/s for YIG, and use air as an example for the dielectric with $\epsilon_{r}= 1$ and $h = 0.1 \lambda_\mathrm{m}$. Figure \ref{fig1}(b) shows the dispersion diagram for $H_1=H_2=893$ G, which is equivalent to $\omega_{0}=\omega'_{0}=0.5\omega_m$. Due to the coupling between SMPs along the two YIG-air interfaces, two nonreciprocal modes (OS and ES) emerge, denoted by the solid and dashed lines. Clearly, a topological unidirectional propagation band occurs in $[\omega_{\rm{sp1}},\omega_{\rm{sp2}}]$, corresponding to $[\omega_{m},1.5\omega_{m}]$, as marked by the yellow shaded area. The dashed lines represent light line with $\omega=\pm kc$. Such unidirectional modes in the bandgap of the YIG bulk modes with $k^2$<$\mu_{\rm{v}}\epsilon_{m}k^2_{0}$ (the gray shaded areas) are topologically protected due to the nontrivial bandgap \cite{mann2021PRL, lu2018nNC}. Moreover, the 2D structure can be accurately extended to a realistic 3D structure with a waveguide width $W$, truncated by two metal slabs along the $z$ direction. To illustrate this, we also numerically solve the modes for the realistic 3D system with modal analysis using COMSOL Multiphysics in Fig. \ref{fig1}(b), and the obtained results  for $W=0.05 \lambda_m$ (see circles) are in good agreement with those for the 2D system. When both the unidirectional ES and OS modes are excited in the same waveguide, USMMI will occur. To verify this, we simulate the wave propagation in the 3D waveguide shown in Fig. \ref{fig1}(c). A line current source with $\omega = 1.1 \omega_m$ is placed at the bottom of the air layer to excite the two unidirectional modes. As expected, the excited wave can only propagate in one direction without any backscattering. Importantly, USMMI with periodic fields of OS and ES is achieved, which can be characterized by the beat length $L_\pi$ \cite{soldano1995JLT}:
	\begin{equation}
		L_\pi=\dfrac{\pi}{\vert{k_{odd}-k_{even}}\vert}
		\label{eq:eq4}
	\end{equation}
	where $k_{odd}$ and $k_{even}$ are the propagation constants of the odd and even modes, respectively. Figure \ref{fig1}(d1) shows the corresponding mode profiles (see the inset) and $E_x$ distributions along the y-axis for OS and ES modes in the 3D system, demonstrating their symmetric features. It should be noted that our interest in this work focuses on the unidirectional region. 
	
	\begin{figure}[t!]
		\centering\includegraphics[width=2.8 in]{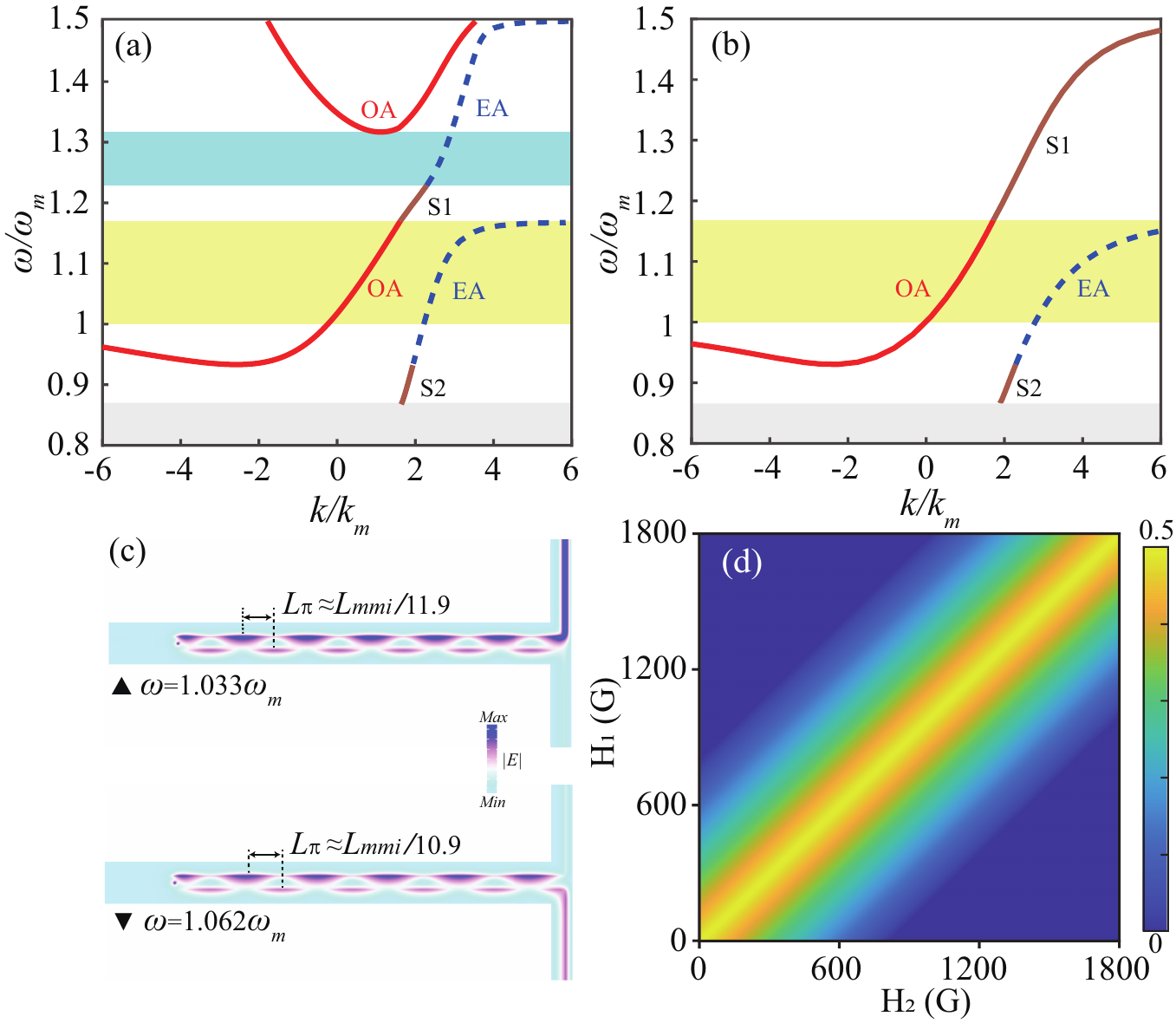}
		\caption{(a, b) Dispersion relation of SMPs in an asymmetric waveguide for $H_1=893$ G and $H_2=300$ G. (a) $d=0.1\lambda_m$  and (b) $d=0.05\lambda_m$. The yellow shaded area represents the region of USMMI between OA and EA double modes, while the bluish shaded area represents the unidirectional EA single mode. S1 and S2 represent the single modes supported at the YIG-air surfaces. (c) Simulated $E$-field amplitudes in the asymmetric splitter at $\omega =1.033 \omega_{m}$ and  $1.062\omega_{m}$. (d) The USMMI bandwidth $\Delta\omega$ as a function of $H_1$ and $H_2$.}\label{fig3}
	\end{figure}
	MMI based on PhCs is useful for designing a tunable splitter \cite{skirlo2014PRL,tang2024LSA}, and either nonlinear mechanisms \cite{Pantazopoulos2019NJP} or, as here, SMP-based MMI can be employed for the same purpose. To verify this, a tunable splitter based on SMP waveguide is proposed in Fig. \ref{fig2}(a). The input waveguide supports USMMI, while the output waveguide supports a single-mode SMP, whose dispersion relation in the metal-YIG-air-metal structure is the same as that of the OS mode in \eqref{eq:eq3b} \cite{hong2021OME,li2024NC}. The point source is placed at a distance of $L_{mmi}$ (the length of the MMI) from the junction. For USMMI, the inverted and direct images of the input field periodically alternate with a constant $L_\pi$, as shown in Fig. \ref{fig1}(c). Figure \ref{fig2}(b) shows the analytic beat length $L_\pi$ using Eq. (\ref{eq:eq4}) as a function of $\omega$, indicated by the red solid line. It can be seen that $L_\pi$ increases with $\omega$ across the entire USMMI region. Moreover, we calculate the numerical values of $L_\pi$ by full-wave simulations for various frequencies, as shown by the circles in Fig. \ref{fig2}(b), which agree well with the analytical values. To verify the tunability of the splitter, the transmission coefficients of the symmetric splitter ($H_1$ = $H_2$ = 893 G) as a function of $\omega$ are shown in Fig. \ref{fig2}(d). Here, we take the loss with $\alpha$=$3\times10^{-5}$ as an example (impact of loss see the Supplement 1). As $\omega$ changes from $1.09 \omega_m$ to $1.15 \omega_m$, the transmission of each output oscillates between nearly $0$ to $1$. The total transmission is always $1$ for lossless ($\alpha=0$) due to the topological unidirectional feature. To clearly illustrate this, the simulated $E$-field amplitudes of the splitter at $\omega = 1.105 \omega_{m}$ and $1.128 \omega_{m}$ are displayed in Fig. \ref{fig2}(c). The value of $L_{mmi}$ satisfies as $L_{mmi}\approx12.1L_{\pi(1.105\omega_p)}\approx11.1L_{\pi(1.128\omega_p)}$, with an inverted (direct) image of the incident field are realized at the upper (lower) corner. Consequently, the unidirectional SMP propagates upward (downward) along Output1 (Output2) at frequencies of \(1.105 \, \omega_m\) (\(1.128 \, \omega_m\)) as expected. The results demonstrate that a frequency splitter based on USMMI is achieved. It should be noted that a magnetically controllable power splitter (see the Supplement 1) can also be realized using USMMI.

	\begin{figure}[t!]
		\centering\includegraphics[width=2.9 in]{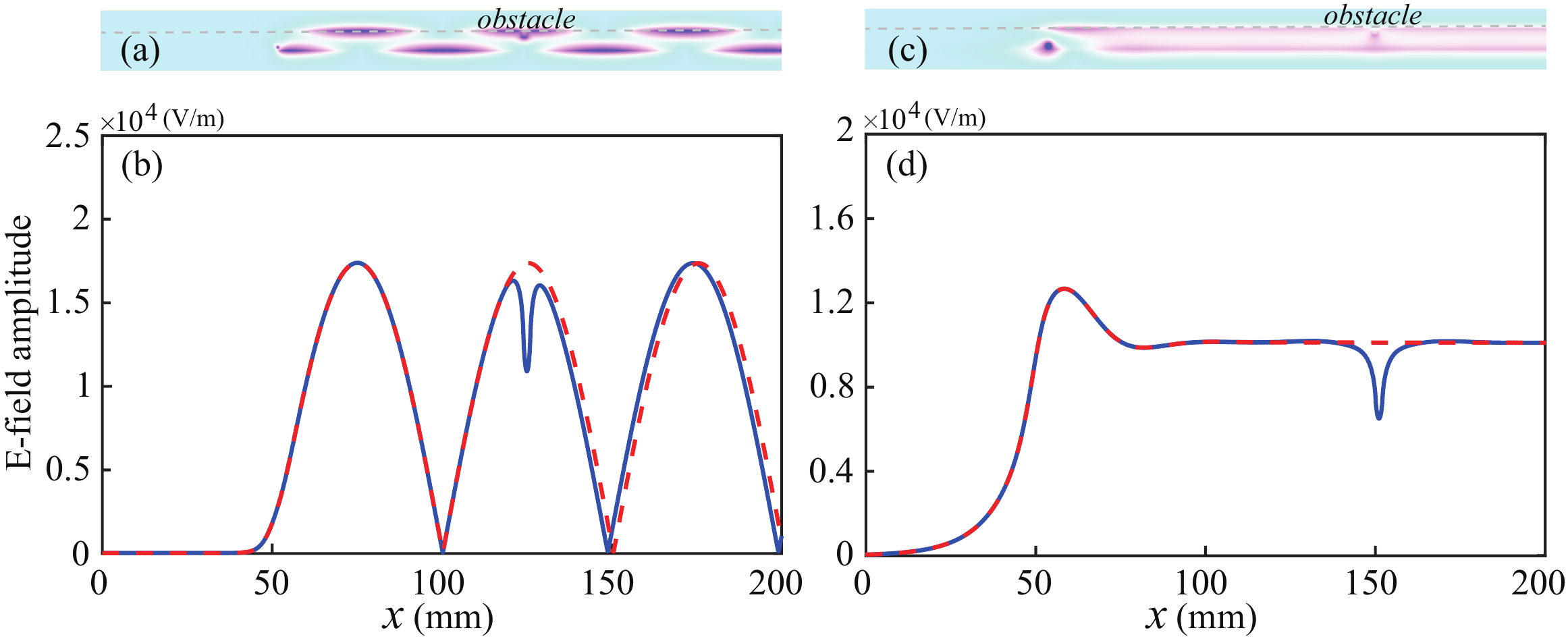}
		\caption{(a, c) Simulated E-field amplitudes in symmetric (a) and asymmetric (c) structures. (b), (d) Distributions of $E$-field amplitudes in (a) and (c) along the upper YIG-air interface (gray dashed lines), respectively. The blue solid and red dashed lines represent the results with and without the obstacles, respectively. The operating frequency is $\omega=1.28\omega_m$.}\label{fig4}
	\end{figure}
	
	Second, we analyze an asymmetric structure ($H_1 \neq H_2$). Here, we take $H_1 = 893$ G and $H_2 = 300$ G as an example, with other parameters being the same as in Fig. (\ref{fig1}). Using \eqref{eq:eq2}, we numerically calculate the dispersion relation of SMPs for the asymmetric waveguide. Figure \ref{fig3}(a) displays the dispersion diagram for $d=0.1\lambda_m$. Due to the asymmetric coupling between modes along the two YIG-air interfaces, the waveguide supports four modes: EA, OA, S1, and S2 modes. The EA and OA modes profiles at $\omega=1.1\omega_m$ are illustrated in Fig. \ref{fig1}(d2), exhibiting even-asymmetric (EA) and odd-asymmetric (OA) characteristics, respectively. The S1 and S2 modes can only propagate at a single surface of the upper or lower YIG-air. As shown in Fig. \ref{fig3}(a), there is also a USMMI band (yellow shaded area) for EA and OA modes in $[\omega_{\rm{sp1}},\omega_{\rm{sp4}}]$, where $\omega_{\rm{sp1}}=\omega_m$ and $\omega_{\rm{sp4}}=1.168\omega_m$. More importantly, there is a bandwidth of $[1.228 \omega_m, 1.315 \omega_m]$ that supports only a single unidirectional EA mode, which differs from the symmetric waveguide shown in Fig. \ref{fig1}(b). The existence of such a single EA mode is due to the strong coupling between the S1 mode and the higher-order EA modes. Figure \ref{fig3}(b) shows the dispersion diagram for $d = 0.05 \lambda_m$. It is found that the band of the single EA mode is significantly affected by the YIG thickness $d$ and disappears when $d$ decreases from $0.1 \lambda_m$ to $0.05 \lambda_m$. Figure \ref{fig3}(c) shows the simulated E-field amplitudes in the asymmetric splitter at $\omega = 1.033 \omega_m$ and $1.062 \omega_m$. Similar to the symmetric splitter, the distance $L_{mmi}$ satisfies $L_{mmi} \approx 11.9 L_{\pi (1.033 \omega_m)} \approx 10.9 L_{\pi (1.062 \omega_m)}$, thus the SMP propagates upward and downward as expected. Moreover, the USMMI bandwidth is not affected by $d$ but is only related to $H_1$ and $H_2$, resulting from the magnetically controllable asymptotic frequency $\omega_{\rm{sp}}$. Figure \ref{fig3}(d) shows the USMMI bandwidth versus the magnetic fields $H_1$ and $H_2$, defined by $\Delta\omega = \max\left(0.5\omega_m - \frac{|H_1 - H_2|}{1786}\omega_m, \, 0\right)$. It can be seen that the bandwidth $\Delta\omega$ is magnetically controllable by varying $H_1$ and $H_2$, and reaching a maximum value of $0.5\omega_m$ when $H_1 = H_2$.

	%\subsection{Topological one-way propagation}
	Due to the topological protection of the unidirectional mode \cite{jinT2016NC}, our proposed SMP waveguides are robust against disorder. To verify this robustness, two 1 mm square YIG obstacles were introduced into the air layer of both the symmetric [Fig. (\ref{fig1})] and asymmetric [Fig. (\ref{fig3})] waveguides. Figures \ref{fig4}(a) and \ref{fig4}(c) show the simulated results of full-wave simulations at $\omega = 1.28 \omega_m$, respectively. As seen in Fig. \ref{fig4}(a), the pattern of USMMI in symmetric waveguide remains almost constant before and after the obstacle. Similarly, it is invariant in the asymmetric waveguide. More importanly, it is found from Fig. \ref{fig4}(c) that the newly emerged unidirectional EA mode effectively circumvents the obstacle without any backscattering, clearly demonstrating its unidirectional dispersion property in Fig. \ref{fig3}(a). Figure \ref{fig4}(b) and \ref{fig4}(d) show the distributions of $E$-field amplitudes along the upper YIG-air interface, corresponding the gray dashed lines in Fig. \ref{fig4}. For comparison, the results without defects are also shown by the  red dashed lines. The field amplitudes closely resemble those with obstacles (blue solid lines), demonstrating the strong robustness of the SMP modes in our proposed symmetric and asymmetric systems.

	\begin{figure}[t!]
		\centering\includegraphics[width=2.8 in]{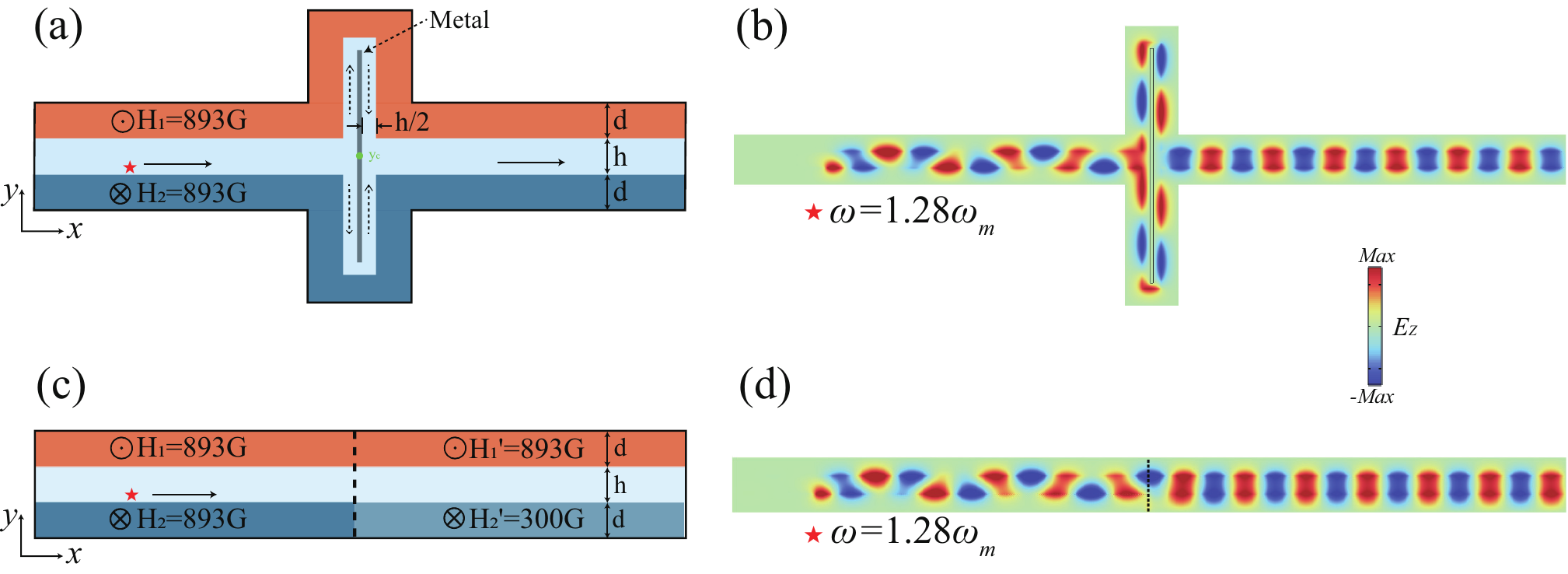}
		\caption{Structure of mode conversion with (a) and without (c) additional metal plate. (b) and (d) Simulated $E_z$ field amplitudes in symmetric (a) and asymmetric (c) structures, demonstrating that the unidirectional multiple modes are transferred to single even mode. The stars mark the source with $\omega=1.28\omega_m$, and the dashed lines in (c,d) represent the boundaries between distinct waveguides.}\label{fig5}
	\end{figure}
	
	%\subsection{Mode conversion}
	
	Finally, we demonstrate the capability of mode conversion between different modes. For this purpose, two different types of mode conversion are considered. The first involves by inserting a metal plate into the waveguide, analogous to the combination of two splitters, as shown in Fig. \ref{fig5}(a). In this waveguide, an excited wave is equally split into two waves, with the difference of initial phase $\Delta{\phi_0}$ and displacement $\Delta{l}$, and then they coupled to a wave. Mode conversion occurs only when $k \Delta{l} + \Delta{\phi_0} = n\pi$, where $\Delta{l} = 4 y_c$, and $y_c$ is the center position of the metal plate along the y-axis.  The metal is assumed to be a perfect electric conductor (PEC) with a length of $1.4\lambda_m$. By appropriately adjusting $y_c$, the incident mode can be converted to an even mode when $n = 0, \pm 2, \pm 4 \ldots$ , and to odd modes when $n = \pm 1, \pm 3, \pm5 \ldots$. Figure \ref{fig5}(b) shows the simulated $E_z$ field pattern for $\omega = 1.28 \omega_m$. It is found that the conversion between multiple modes and even mode can be achieved, when $y_c=-0.036\lambda_m$. Furthermore, the conversion between the even and odd modes can also be realized by varying $y_c$ in this waveguide. More importantly, the second type of mode conversion, without additional metal plate to change the wave phase, is proposed by connecting the two waveguides shown in Fig. \ref{fig5}(c). In this structure, the left part is a symmetric waveguide [Fig. \ref{fig4}(a)], while the right part is an asymmetric waveguide [Fig. \ref{fig4}(b)]. Figure \ref{fig5}(d) shows the simulated $E_z$ field pattern for $\omega=1.28 \omega_m$. Since only one even mode exists in the right waveguide at this frequency, the excited multiple modes are converted into a single even mode as expected, possessing the advantages of simple mode conversion structure. Therefore, we conclude that mode conversion, both with and without the insertion of metal, can be achieved.

	%\section{Conclusion}
		In conclusion, we have proposed a waveguide composed of two YIG slabs sandwiched between metal and dielectric layers, which supports multiple SMP modes. The dispersion properties of these SMPs have been analyzed, exhibiting unidirectional feature. We demonstrated that robust USMMI can be achieved in SMP waveguides, overcoming the limitation of backscattering in traditional waveguides. Furthermore, tunable splitters based on USMMI have been designed in both symmetric and asymmetric structures. USMMI has been shown to be immune to disorders, and mode conversion can also be realized. Notably, the asymmetric waveguide supports only an even mode within a specific single-mode frequency range, differing from the behavior observed in symmetric waveguide. These results can be extended to terahertz and optical frequencies, offering significant flexibility to manipulate topological waves.

	\bigskip
	\begin{backmatter}
		\bmsection{Funding} National Natural Science Foundation of China (12464057, 12104203); Jiangxi Provincial Natural Science Foundation (20242BAB25039, 20224BAB211015).
		
		\bmsection{Disclosures} The authors declare no conflicts of interest.
		
		\bmsection{Data Availability Statement} The data supporting the findings of this study are available from the corresponding author upon reasonable request.
		
	\bmsection{Supplemental document}See Supplement 1 for supporting content.
	\end{backmatter}
	
	% Bibliography
	\bibliography{sample}

	\ifthenelse{\equal{\journalref}{aop}}{%
		\section*{Author Biographies}
		\begingroup
		\setlength\intextsep{0pt}
		\begin{minipage}[t][6.3cm][t]{1.0\textwidth} % Adjust height [6.3cm] as required for separation of bio photos.
			\begin{wrapfigure}{L}{0.25\textwidth}
				\includegraphics[width=0.25\textwidth]{john_smith.eps}
			\end{wrapfigure}
			\noindent
			{\bfseries John Smith} received his BSc (Mathematics) in 2000 from The University of Maryland. His research interests include lasers and optics.
		\end{minipage}
		\begin{minipage}{1.0\textwidth}
			\begin{wrapfigure}{L}{0.25\textwidth}
				\includegraphics[width=0.25\textwidth]{alice_smith.eps}
			\end{wrapfigure}
			\noindent
			{\bfseries Alice Smith} also received her BSc (Mathematics) in 2000 from The University of Maryland. Her research interests also include lasers and optics.
		\end{minipage}
		\endgroup
	}{}

\end{document}

\maketitle

\section*{Supplementary Material 1: Derivation of the analytic dispersion}

Here, we solve the dispersion relation of the SMP in a metal-YIG-dielectric-YIG-metal structure under two opposite static magnetic fields, as shown in Fig. (1a) of the main text. This waveguide only supports the TE mode ($E_x$ = $E_y$ = $H_z$ = 0), which satisfies the Maxwell's equations: $\nabla \times \bm{E} = i \omega \mu_0 \bm{\mu}^\pm_m \bm{H}$ and $\nabla \times \bm{H} = -i\omega \varepsilon \varepsilon_0 \bm{E}$. By substituting $\bm{\mu}^\pm_m$ (Eq. 1 in the main text), we can write all the scalar equations as 
\begin{align}
	&\dfrac{\partial E_{z}}{\partial y} = i\omega\mu_0 (\mu_1 H_x + i\mu_2 H_y) \label{S1}\\[6pt]
	-&\dfrac{\partial E_{z}}{\partial x} = i\omega\mu_0 (\mu_1 H_y - i\mu_2 H_x) \label{S2}\\[6pt]
	&\dfrac{\partial H_y}{\partial x} - \dfrac{\partial H_x}{\partial y} = -i\omega\varepsilon_0\varepsilon E_z \label{S3}
\end{align}
in the upper YIG layer, while the equations for the lower YIG layer can be obtained by replacing $\mu_1'$ and $\mu_2'$ with $\mu_1$ and $\mu_2$ in the Eq. (S1) and (S2). In the dielectric layer, $\mu_1=1$ and $\mu_2=0$. Considering plane waves, the electric field component $E_{z}$ is expressed as 
\begin{align}
	\bm{E_{z1}} & = (A_1 e^{-{\alpha_1}y} + A_2 e^{{\alpha_1}y}) e^{i(kx - \omega t)}, \quad y \in \scriptstyle{\left[ \frac{h}{2}, \frac{h}{2} + d \right)} \nonumber \\[6pt]
	\bm{E_{z2}} & = (B_1 e^{-\alpha_ry} + B_2 e^{\alpha_ry}) e^{i(kx - \omega t)}, \quad y \in \scriptstyle{\left[ -\frac{h}{2}, \frac{h}{2} \right)} \label{S4}\\[6pt]
	\bm{E_{z3}} & = (C_1 e^{-\alpha_2y} + C_2 e^{\alpha_2y}) e^{i(kx - \omega t)}, \quad y \in \scriptstyle{\left[ -\frac{h}{2} - d, -\frac{h}{2} \right)} \nonumber
\end{align}
for the upper YIG layer, middle dielectric layer and lower YIG layer, respectively, where  $A_1$, $A_2$, $B_1$, $B_2$, $C_1$ and $C_2$ are the amplitude of the ﬁeld. Note that the attenuation coefficients (${\alpha_1}$, ${\alpha_r}$, ${\alpha_2}$) and the other parameters (such as: $k$, $\mu_v$, $\mu_v'$) are presented in the main text. 

By combining Eq. (S1) and (S2), the nonzero components ($H_x$, $H_y$) of the magnetic field can be directly derived from $E_z$, thus $H_x$ for three layers can be obtained by substituting $E_{z1}$, $E_{z2}$ and $E_{z3}$ as
\begin{align}
	&\bm{H_{x1}} =\dfrac{i}{\mu_v\mu_0 \omega} \left[(k\dfrac{\mu_2}{\mu_1}+\alpha_1)A_1 e^{-\alpha_1 y} + (k\dfrac{\mu_2}{\mu_1} - \alpha_1)A_2 e^{\alpha_1 y} \right] e^{i(kx - \omega t)}\nonumber\\[8pt]
	&\bm{H_{x2}} =\dfrac{i\alpha_r}{\mu_0 \omega}(B_1e^{-\alpha_r y} - B_2e^{\alpha_r y})e^{i(kx - \omega t)} \label{S5}\\[6pt]
	&\bm{H_{x3}} =\dfrac{i}{\mu_v'\mu_0 \omega} \left[(-k\dfrac{\mu'_2}{\mu'_1}+\alpha_2)C_1e^{-\alpha_2 y} - (k\dfrac{\mu_2'}{\mu_1'} + \alpha_2)C_2e^{\alpha_2 y}\right]e^{i(kx - \omega t)}\nonumber
\end{align}

According to the boundary conditions of electric and magnetic fields, $E_z$ and $H_x$ are continuous at the YIG-dielectric interfaces $y = \pm\frac{h}{2}$, while the electric field $E_z$ is zero at the YIG-metal interfaces $y = 0$, where the metal is assumed to be a perfect electric conductor (PEC). Therefore, we obtain six boundary equations as follows
\begin{equation}
	\begin{cases}
		\text{\ding{172}}\quad E_{z1}\mid_{y=\frac{h}{2}+d} = 0 \quad (\text{PEC boundary})
		\\[6pt]
		\text{\ding{173}}\quad E_{z1}\mid_{y=\frac{h}{2}} = E_{z2}\mid_{y=\frac{h}{2}}
		\\[6pt]
		\text{\ding{174}}\quad H_{x1}\mid_{y=\frac{h}{2}} = H_{x2}\mid_{y=\frac{h}{2}}
		\\[6pt]
		\text{\ding{175}}\quad E_{z2}\mid_{y=-\frac{h}{2}} = E_{z3}\mid_{y=-\frac{h}{2}}
		\\[6pt]
		\text{\ding{176}}\quad H_{x2}\mid_{y=-\frac{h}{2}} = H_{x3}\mid_{y=-\frac{h}{2}}
		\\[6pt]
		\text{\ding{177}}\quad E_{z3}\mid_{y=-\frac{h}{2}-d} = 0 \quad (\text{PEC boundary})
		\label{S6}
	\end{cases}
\end{equation}

By substituting \( E_{z} \) and \( H_{x} \) from Eqs. (S4) and (S5) into Eqs. (S6), we obtain
\begin{equation}
	\begin{cases}
		\text{\ding{172}}\quad  A_1 e^{-\alpha_1(\frac{h}{2}+d)} + A_2 e^{\alpha_1(\frac{h}{2}+d)} = 0
		\\[6pt]
		\text{\ding{173}}\quad A_1 e^{-\alpha_1\frac{h}{2}} + A_2 e^{\alpha_1 \frac{h}{2}} = B_1 e^{-\alpha_1 \frac{h}{2}} +   B_2 e^{\alpha_1 \frac{h}{2}}
		\\[6pt]
		\text{\ding{174}}\quad \dfrac{1}{\mu_v} \left[(k\dfrac{\mu_2}{\mu_1}+\alpha_1)A_1 e^{-\alpha_1 \frac{h}{2}} + (k\dfrac{\mu_2}{\mu_1} - \alpha_1)A_2e^{\alpha_1 \frac{h}{2}}\right] = \alpha_r(B_1e^{-\alpha_r \frac{h}{2}} - B_2e^{\alpha_r \frac{h}{2}})
		\\[6pt]
		\text{\ding{175}}\quad B_1 e^{\alpha_r \frac{h}{2}} + B_2 e^{-\alpha_r\frac{h}{2}} = C_1 e^{\alpha_2 \frac{h}{2}} + C_2 e^{-\alpha_2 \frac{h}{2}}
		\\[6pt]
		\text{\ding{176}}\quad \dfrac{1}{\mu_v'} \left[(-k\dfrac{\mu'_2}{\mu'_1}+\alpha_2)C_1e^{\alpha_2 \frac{h}{2}} - (k\dfrac{\mu_2'}{\mu_1'} + \alpha_2)C_2e^{-\alpha_2 \frac{h}{2}}\right] = \alpha_r(B_1e^{\alpha_r \frac{h}{2}} - B_2e^{-\alpha_r \frac{h}{2}})
		\\[6pt]
		\text{\ding{177}}\quad C_2 + C_1e^{2\alpha_2 (\frac{h}{2}+d)} = 0
		\label{S7}
	\end{cases}
\end{equation}

Finally, by eliminate the coefficients \( A_1, A_2, B_1, B_2, C_1, C_2 \) from the six boundary equations in Eqs. (S7), we can solve for the dispersion relation of the SMP
\begin{equation}
	e^{2\alpha_r h} = \dfrac{\left[1 - \dfrac{1}{\alpha_r\mu_v}(k\dfrac{\mu_2}{\mu_1} + \dfrac{\alpha_1}{\tanh(\alpha_1d)})\right]\left[1 - \dfrac{1}{\alpha_r\mu_v'}(k\dfrac{\mu_2'}{\mu_1'} + \dfrac{\alpha_2}{\tanh(\alpha_2d)})\right]}{\left[1 + \dfrac{1}{\alpha_r\mu_v}(k\dfrac{\mu_2}{\mu_1} + \dfrac{\alpha_1}{\tanh(\alpha_1d)})\right]\left[1 + \dfrac{1}{\alpha_r\mu_v'}(k\dfrac{\mu_2'}{\mu_1'} + \dfrac{\alpha_2}{\tanh(\alpha_2d)})\right]}.
	\label{S8}
\end{equation}

Simplifying the dispersion equation, Eq. (S8) becomes
\begin{equation}
	e^{2\alpha_r h}=\frac{\left(1-\dfrac{M}{\alpha_r\mu_v}\right)
		\left(1-\dfrac{N}{\alpha_r\mu_v'}\right)}
	{\left(1+\dfrac{M}{\alpha_r\mu_v}\right)
		\left(1+\dfrac{N}{\alpha_r\mu_v'}\right)}
	\label{S9}
\end{equation}
with  $\bm{M}= k\dfrac{\mu_2}{\mu_1} + \dfrac{\alpha_1}{\tanh{\alpha_1d}}$ and $\bm{N} = k\dfrac{\mu_2'}{\mu_1'} + \dfrac{\alpha_2}{\tanh{\alpha_2d}}$, which corresponding Eq. 2 in the main text.

When considering a symmetry structure where $H_1=H_2$, we have $\alpha_1 = \alpha_2$, $\mu_2 = \mu_2'$, $\mu_1 = \mu_1'$, and $\mu_v = \mu_v'$. Thus, $M=N$ in Eq. (S9), and we obtain
\begin{equation}
	e^{2\alpha_r h} = \left( \frac{1 - \dfrac{M}{\alpha_r \mu_v}}
	{1 + \dfrac{M}{\alpha_r \mu_v}} \right)^2
	\label{S10}
\end{equation}

Obviously, there are two solutions in Eq. (S10). Combining these solutions with the formula of $tanh(x)=\dfrac{e^{x}-e^{-x}}{e^{x}+e^{-x}}$, we have
\begin{equation}
	e^{\alpha_r h} = \frac{1 + \tanh\left(\frac{\alpha_r h}{2}\right)}{1 - \tanh\left(\frac{\alpha_r h}{2}\right)}
	= \pm \frac{1 - \dfrac{M}{\alpha_r \mu_v}}
	{1 + \dfrac{M}{\alpha_r \mu_v}} 
	\label{S11}
\end{equation}

From Eq.(S10), the dispersion relation of SMP can be simplified
\begin{subequations}
	\begin{eqnarray}
		k\dfrac{\mu_2}{\mu_1} + \dfrac{\alpha_1}{\tanh{\alpha_1d}} +\alpha _{r}\mu_{\rm{v}}{\rm{tanh}}\left (\dfrac{\alpha _{r}h}{2} \right )=0 \quad(ES) \label{S12a}\\[8pt]
		k\dfrac{\mu_2}{\mu_1} + \dfrac{\alpha_1}{\tanh{\alpha_1d}}+
		\alpha _{r}\mu_{\rm{v}}{\rm{coth}}\left (\dfrac{\alpha _{r}h}{2} \right )=0 \label{S12b}\quad(OS)
	\end{eqnarray}
\end{subequations}
for the even-symmetric (ES) and odd-symmetric (OS) modes, respectively, which correspond Eq. 3 in the main text.

\section*{Supplementary Material 2: the waveguide properties for different parameters}

In the main text, we use fixed values for parameters (such as loss and magnetic field) as examples to investigate the performance of the waveguide. Here, we analyze the impact of varying these parameter values on the waveguide characteristics.

\begin{figure}[H]
	\centering\includegraphics[width=4.7 in]{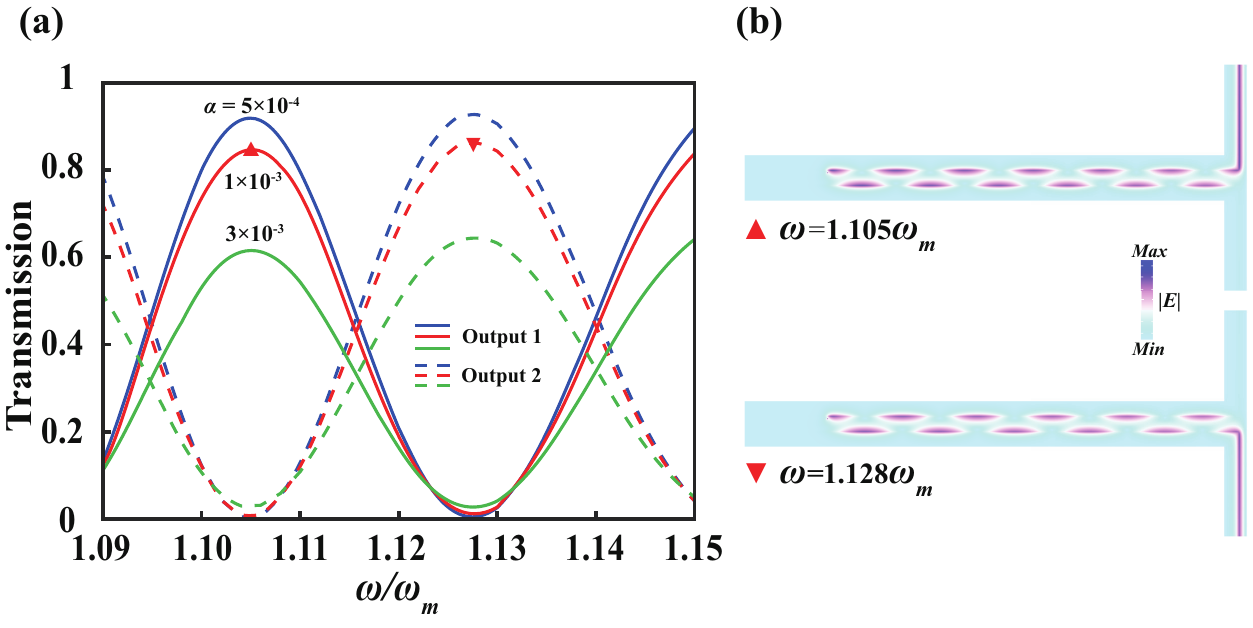}
	\caption{The effect of YIG loss $\alpha$ on the USMMI-based splitter. (a) Transmission coefficients of the symmetric splitter as a function of $\omega$ for different values of $\alpha$. (b) Simulated $E$-field amplitude for $\alpha=1\times10^{-3}$ at $\omega=1.105\omega_m$ and $\omega=1.128\omega_m$. The other parameters are the same as in Fig. 2 of the main text.}
	\label{figS1}
\end{figure}

We first investigate the impact of YIG loss on the USMMI-based splitter. Given that the YIG material used in different experiments exhibits varying loss coefficients \cite{tang2024LSA,tong2016APL}, we analyze the transmission coefficients of the symmetric splitter ($H_1 = H_2 = 893$ G) under different YIG loss coefficients, as shown in Fig. \ref{figS1}(a). As the loss varies from $\alpha = 3\times10^{-5}$ to $\alpha = 3\times10^{-3}$, the splitting ratio of each output can be tuned by adjusting the frequencies, while the total transmission decreases with $\alpha$. Fig. \ref{figS1}(b) shows the simulated E-field amplitudes for $\alpha=1\times10^{-3}$. Similar to the results for $\alpha=3\times10^{-5}$ shown in Fig. 2(d) of the main text, the unidirectional SMP propagates upward and downward at $\omega = 1.105 \omega_{m}$ and $1.128 \omega_{m}$, as expected. Therefore, we conclude that the frequency splitter based on USMMI can be achieved for different YIG loss values. It should be noted that for a larger loss of $\alpha= 3\times10^{-3}$, non-zero power is clearly observed at $\omega = 1.105 \omega_{m}$, while the power in the downward port is nearly zero for $\alpha= 1\times10^{-3}$, as indicated by the dashed line in Fig. \ref{figS1}(a). To further investigate this phenomenon, we independently excite the odd and even modes and calculate their transmission efficiencies ($\eta_{odd}$ and $\eta_{even}$) over a distance of $\lambda_{m}$. Here, we take $\omega=1.1\omega_{m}$ as an example. We find that $\eta_{odd}=99.6\%$ and $\eta_{even}=99.9\%$ for $\alpha = 3\times10^{-5}$, $\eta_{odd}=92.9\%$ and $\eta_{even}=98.1\%$ for $\alpha = 5\times10^{-4}$, $\eta_{odd}=86.3\%$ and $\eta_{even}=96.2\%$ for $\alpha = 1\times10^{-3}$, and $\eta_{odd}=64.2\%$ and $\eta_{even}=89.0\%$ for $\alpha = 3\times10^{-3}$. The results indicate that non-zero power arises from the significant difference in transmission losses between the odd and even modes at higher YIG loss values, while the difference in transmission losses remains smallish for reasonable YIG losses, enabling near 0 to 1 splitting performance.
	
  \begin{figure}[H]
  	\centering\includegraphics[width=5 in]{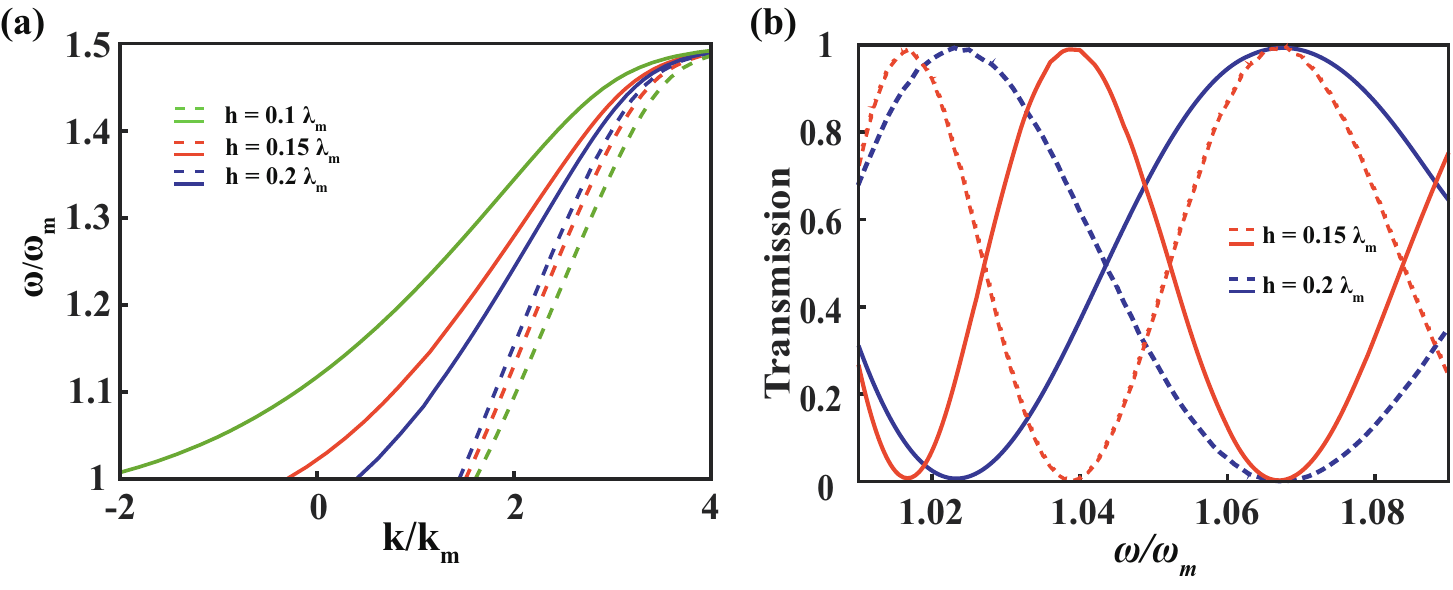}
  	\caption{The impact of the dielectric width $h$. (a) The dispersions of odd (solid lines) and even (dashed lines) SMP modes vary with different $h$ in the whole USMMI band. (b) Transmission coefficients of the symmetric splitter as a function of $\omega$ for two different values of $h$: $0.15\lambda_{m}$ and $0.2\lambda_{m}$, where the solid and dashed lines represent the transmission in the upper and lower ports of the splitter.}
  	\label{figS2}
  \end{figure}
  
   The impact of the dielectric width $h$ on the properties of the waveguide is further explored. Figure \ref{figS2}(a) shows the calculated dispersion curves for the SMP at various $h$ values using MATLAB software. It can be observed that as $h$ increases from $h=0.1\lambda_{m}$ to $h=0.2\lambda_{m}$, the odd and even modes supported by the SMP waveguide gradually converge; however, the unidirectional propagation bands remain consistent within the range of [$\omega_m$, $1.5\omega_m$]. By maintaining all other parameters consistent with those in Fig. 2(d) of the main text, we calculated the transmission coefficients for different $h$ values. As shown in Fig. \ref{figS2}(b), the beam-splitting performance remains nearly between 0 and 1 despite changes in $h$, when the frequency changes from $1.01\omega_m$ to $1.09\omega_m$. Due to variations in $h$, the beat length $L_\pi$ at the same frequency changes, resulting in different splitting ratios at the same frequency. The results demonstrate that a frequency-tunable splitter can be achieved for different dielectric thicknesses.

    \begin{figure}[H]
    	\centering\includegraphics[width=4.8 in]{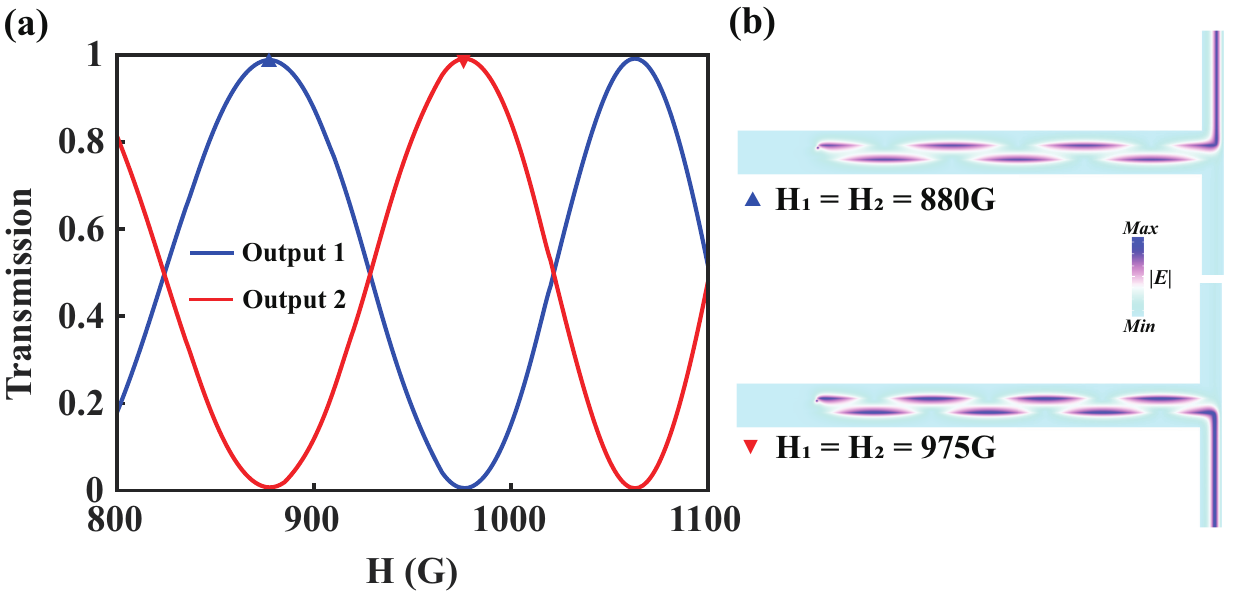}
    	\caption{Magnetically controllable splitter. (a) Transmission coefficient as a function of the magnetic field for $H_1 = H_2$. (b) Simulated $E$-field amplitude for $H_1 = H_2 = 880\, \text{G}$ and $H_1 = H_2 = 975 \, \text{G}$, clearly demonstrating that the energy is almost entirely directed to the upper port (output 1) and the lower port (output 2), respectively. The working frequency is $\omega=1.1\omega_m$. }
    	\label{figS3}
    \end{figure}
    
	It should be noted that, for our proposed waveguide, a magnetically tunable power splitter based on USMMI can be realized at a fixed frequency. To verify this, we calculate the transmission coefficients of the symmetric splitter ($H_1 = H_2$) as a function of the external magnetic field $H$, as shown in Fig. \ref{figS3}(a). Here, we choose a fixed frequency of $\omega=1.1\omega_m$ as an example. It is clearly shown that the beam splitting ratio can be achieved from 0 to 1 by changing the $H$ value from $880$ G to $975$ G. Figure \ref{figS3}(b) shows the simulated $E$-field amplitude for $H_1 = H_2 = 880\, \text{G}$ and $H_1 = H_2 = 975 \, \text{G}$, clearly demonstrating that the energy is almost entirely directed to the upper port (output 1) and the lower port (output 2), respectively. Therefore, a magnetically controllable power splitter is realized using USMMI.

\begin{figure}[t]
	\centering\includegraphics[width=5 in]{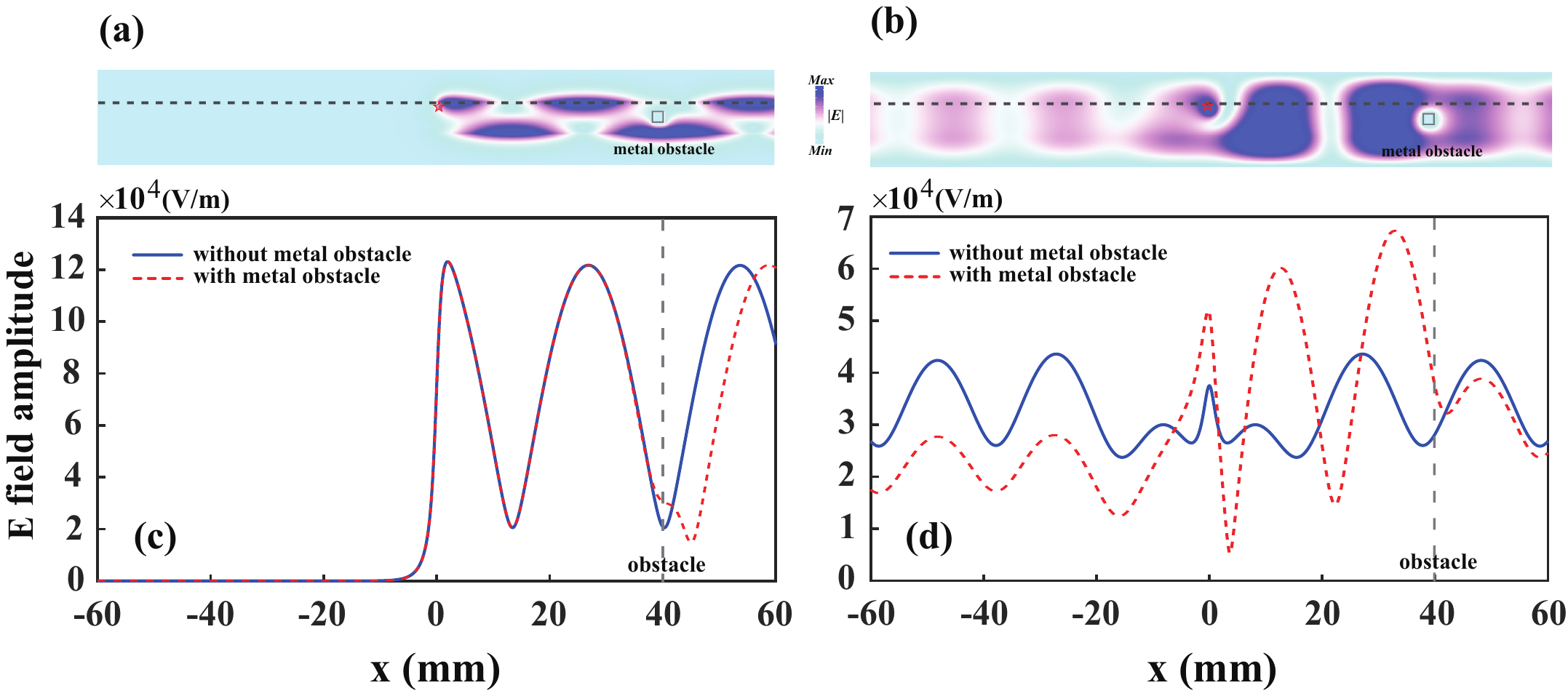}
	\caption{Comparison of the waveguide with (a, c) and without (b, d) external magnetic field. E-field distribution diagrams. (a) A unidirectional mode without backscattering. (b) A regular mode with strong backscattering. (c, d) The field distribution with (the red dashed lines) and without (the blue solid lines) a 2 mm square metallic obstacle along the YIG-air interface, indicated by the black dashed lines in (a) and (b). The red star shows the position of the point source. $H_1=H_2=893$ G, and $\omega=1.1\omega_m$.
	}
	\label{figS4}
\end{figure}

Finally, we compared our magnetized waveguide with a non-magnetized waveguide, as shown in Fig. S4. To investigate their robustness against defect, we introduced a square metallic obstacle with a length of 2 mm on the right side of the waveguide. Figures \ref{figS4}(a) and \ref{figS4}(b) show the simulated $E$-field results with and without the external magnetic field, respectively. In the full-wave simulations, a point source with $\omega = 1.1 \omega_m$ is used to excite the mode, marked by a red star.
As seen in Fig. \ref{figS4}(a), the excited mode in our magnetized waveguide can only propagate forward, not backward, demonstrating its characteristic of unidirectional propagation without backscattering, even when the obstacle is introduced. In contrast, the mode in the non-magnetized waveguide can propagate both forward and backward, and strong backscattering occurs due to the obstacle, as shown in Fig. \ref{figS4}(b). To clearly illustrate this, Figures \ref{figS4}(c) and \ref{figS4}(d) show the $E$-field distributions along the upper YIG-air interface, with and without obstacles, respectively. As seen from the solid and dashed lines in Fig. \ref{figS4}(c), the field amplitudes on the left side of the obstacle completely overlap with and without the obstacle, and it recovers after passing through the obstacle, demonstrating the strong robustness of our magnetized waveguide. However, the non-magnetized waveguide exhibits a symmetric field distribution without the obstacle due to its reciprocal propagation (see the blue solid line in Fig. \ref{figS4}(d)), and significant changes occur in the distribution due to the strong backscattering from the obstacle (see the red dashed line in Fig. \ref{figS4}(d)). These results further validate the advantage of our structure in achieving robust topological unidirectional propagation without backscattering.

%The supplemental document may contain linked objects such as video, 2D, 3D, and machine-readable data files. Please see the \href{https://opg.optica.org/submit/style/supplementary_materials.cfm}{Author Guidelines for Supplementary Materials} for more information. Such files should be cited in the supplementary document as in the primary document but using the naming convention described above.

%\section*{References} 

%The supplementary materials document may contain a reference list. The reference list should follow our citation style and should be checked carefully, since staff will not be performing any copyediting. You may add citations manually or use BibTeX. See \cite{Zhang:14}.

%Citations that are relevant to the primary manuscript and the supplementary document may be included in both places.

% Bibliography
\bibliography{sample}

%Manual citation list
%\begin{thebibliography}{1}
%\bibitem{Zhang:14}
%Y.~Zhang, S.~Qiao, L.~Sun, Q.~W. Shi, W.~Huang, %L.~Li, and Z.~Yang,
 % \enquote{Photoinduced active terahertz metamaterials with nanostructured
  %vanadium dioxide film deposited by sol-gel method,} Opt. Express \textbf{22},
  %11070--11078 (2014).
%\end{thebibliography}